\documentclass[11pt,english]{article}
\usepackage{geometry}
\usepackage{verbatim}
\usepackage{textcomp}
\usepackage{amsmath}
\usepackage{amssymb}
\usepackage{esint}
\makeatletter

\@ifundefined{definecolor}
 {\usepackage{color}}{}

\usepackage{babel}

\usepackage{graphicx}
\makeatother
\usepackage{hyperref}

\newcommand{\be}{\begin{equation}}
\newcommand{\ee}{\end{equation}}
\newcommand{\bea}{\begin{eqnarray}}
\newcommand{\eea}{\end{eqnarray}}
\newcommand{\beas}{\begin{eqnarray*}}
\newcommand{\eeas}{\end{eqnarray*}}

\def\be{\begin{eqnarray}}
\def\ee{\end{eqnarray}}

\def\>{\rangle}
\def\<{\langle}

\def\ba{\begin{align}}
\def\ea{\end{align}}
\def\bas{\begin{align*}}
\def\eas{\end{align*}}

\newcommand{\htt}{\hat{t}}
\newcommand{\hp}{\hat{\phi}}

\begin{document}
\begin{titlepage} 


\begin{center}
\vspace{2mm}

\par\end{center}

\begin{center}
\textbf{\Large{}Bulk metric reconstruction from boundary entanglement}\textbf{ }
\par\end{center}

\begin{center}
Shubho R. Roy$^{1}$ and Debajyoti Sarkar$^{2}$, 
\\
 
\par\end{center}

\begin{center}
$^{1}$\textsl{\small{}Department of Physics}\textsl{ }\\
\textsl{\small{}Indian Institute of Technology, Hyderabad}\textsl{ }\\
\textsl{\small{}Kandi, Sangareddy, Telengana, India 502285}\textsl{ }\\
\texttt{\textsl{\small{}sroy@iith.ac.in}}~\\

\par\end{center}{\small \par}

%
%
\begin{center}
$^{2}$\textsl{\small{}Albert Einstein Center for Fundamental Physics}\\
\textsl{\small{}Institute for Theoretical Physics}\\
\textsl{\small{}University of Bern, Switzerland}\\
\texttt{\small{} sarkar@itp.unibe.ch }\\
\end{center}
%
%

\vskip 1.5 cm
\begin{abstract}
Most of the literature in the \emph{bulk reconstruction program} in holography focuses on recovering local bulk operators propagating on a quasilocal bulk geometry and the knowledge of the bulk geometry is always assumed or guessed. The fundamental problem of the bulk reconstruction program, which is \emph{recovering the bulk background geometry (metric)} from the boundary CFT state is still outstanding. In this work, we formulate a recipe to extract the bulk metric itself from the boundary state, specifically, the modular Hamiltonian information of spherical subregions in the boundary. Our recipe exploits the recent construction of Kabat and Lifschytz \cite{Kabat:2017mun} to first compute the bulk two point function of scalar fields directly in the CFT without knowledge of the bulk metric or the equations of motion, and then to take a large scaling dimension limit (WKB) to extract the geodesic distance between two close points in the bulk i.e. the metric. As a proof of principle, we consider three dimensional bulk and selected CFT states such as the vacuum and the thermofield double states. We show that they indeed reproduce the pure AdS and the regions outside the Rindler wedge and the BTZ black hole \emph{up to a rigid conformal factor}. Since our approach does not rely on symmetry properties of the CFT state, it can be applied to reconstruct asymptotically AdS geometries dual to arbitrary general CFT states provided the modular Hamiltonian is available. We discuss  several obvious extensions to the case of higher spacetime dimensions as well as some future applications, in particular, for constructing metric beyond the causal wedge of a boundary region. In the process, we also extend the construction of \cite{Kabat:2017mun} to incorporate the first order perturbative locality for AdS scalars. 
\end{abstract}
\end{titlepage}

\setcounter{footnote}{0}
\renewcommand\thefootnote{\mbox{\arabic{footnote}}}

\section{Introduction}\label{sec:intro}
AdS$_{d+1}$/CFT$_d$ duality \cite{Maldacena:1997re,Gubser:1998bc,Witten:1998qj} is an exact equivalence between quantum gravity theories in asymptotically Anti de Sitter (aAdS) spacetimes and a non-gravitational conformal field theory (CFT) living in the conformal boundary of the aAdS spacetime. Thus for the first time, this remarkable duality furnishes a UV-complete and fully non-perturbative description of quantum gravity of any kind in terms of a quantum field theory. Naturally this has opened the doors to understanding the effects of full quantum gravity and various puzzles and paradoxes of general relativity (namely, black hole information paradox, structure of black hole interiors and gravitational singularities etc.), in terms of well-defined and in many cases, controllable field theory calculations. Since the inception of the duality conjecture, efforts have been made to uncover how in general a quasilocal bulk (spacetime and matter) emerges from the underlying CFT state and operator spectrum at large $N$. A precise question that one can ask is how do local quantum fields in the bulk i.e. semiclassical curved space QFT arise from strongly coupled conformal field theory. In the CFT  this corresponds to limit of large $N$ as well as large 't Hooft coupling. A major landmark result in this direction is the ``HKLL smearing function"  construction of \cite{Hamilton:2005ju,Hamilton:2006az,Hamilton:2006fh} which reconstructs local bulk (scalar) fields in terms of non-local (smeared out) boundary CFT primary operators \emph{with compact support}. At the level of free fields, which corresponds to the leading order planar limit of field theory, it was shown that the resulting CFT smeared operators indeed satisfy free bulk field equations in backgrounds such as pure AdS, Rindler, BTZ etc. Since then, it has found several extensions including free higher spin fields in the bulk \cite{Kabat:2012hp,Sarkar:2014dma}, connection with RG flows and dS \cite{Sarkar:2014jia,Xiao:2014uea}, black hole information problem \cite{Lowe:2008ra,Kabat:2014kfa,Roy:2015pga,Kabat:2016rsx} etc. With HKLL it became technically much easier to study the dynamics of quantum fields in the classical bulk spacetime, including their locality properties. It also, very organically, paves the way of incorporating the perturbative $1/N$ or $1/\lambda$ corrections \cite{Roy:2017hcp} by demanding that the bulk commutators obey bulk microcausality order by order in $1/N$ \cite{Kabat:2011rz,Kabat:2012av,Kabat:2013wga,Kabat:2015swa}.\footnote{Very recently, the same considerations also found an illuminating connection with conformal bootstrap \cite{Kabat:2016zzr}.} Thus the HKLL program provides us a bottom-up view of the semiclassical window of quantum gravity theories in terms of correlators of (smeared) operators. However, a huge limitation of the HKLL program is that one necessarily needs the knowledge of the bulk metric and field equations of motion to reconstruct it from the CFT. For the bulk to be truly emergent, the bulk metric and the equations of motion of bulk fields should be \emph{end products}, not ingredients of the bulk reconstruction recipe. This paper aims to extract the bulk spacetime metric itself from the boundary CFT data. This is a highly non-trivial issue since it is not at all clear that, even in principle, whether one should be able to construct the bulk metric given the state and spectrum of operators, and in particular which CFT information is directly related to the bulk geometry. For this we note a parallel line of development in the holographic mapping, since the pioneering work of Ryu and Takayanagi \cite{Ryu:2006bv}, that of the boundary entanglement and the emergent bulk geometry. See e.g \cite{VanRaamsdonk:2009ar,Faulkner:2013ica,Almheiri:2014lwa,Pastawski:2015qua,Jafferis:2015del,Sanches:2017xhn,Faulkner:2017tkh} among many others. Recently \cite{Kabat:2017mun} showed that the boundary entanglement structure can also be used to provide an alternative derivation of the HKLL prescription, which doesn't require the knowledge of the bulk equations of motion or even the bulk metric! Thus, one no longer needs any information about the bulk dynamics in order to construct a bulk scalar field at any order of $1/N$ or $1/\lambda$ expansion. This construction clears the path for us to recover the bulk metric itself as follows. One first computes the supergravity (SUGRA) correlator of two bulk scalars directly in the CFT using \cite{Kabat:2017mun} i.e. without the need for knowing the bulk metric or the bulk equation of motion for the scalar fields. Subsequently, we take the large conformal dimension limit, $\Delta \rightarrow \infty$ of this CFT result, and we identify it with the WKB form of the SUGRA two-point function. From this identification, the bulk metric can be read off since in the WKB limit, the propagator is expected to be exponential of the geodesic length joining the two point. Following this sequence of steps, we show, that starting from the information of boundary modular Hamiltonian one does reproduce the metric, but \emph{up to a conformal factor} which can in general be spacetime dependent. We trace this subtlety back to the field redefinition ambiguity of the bulk fields. However, as we discuss at the end, with an additional intuition regarding the bulk that comes up in the process, it is in fact possible to obtain the bulk metric exactly.  In addition we also show that the prescription of \cite{Kabat:2017mun} can be extended to obtain the bulk scalar, which is local at least up to the first order of bulk perturbations.\\

The plan of the paper is as follows. In section \ref{sec:DGrev} we briefly review the work of \cite{Kabat:2017mun}, which uses the CFT modular Hamiltonian data to recover the HKLL construction of the \emph{free} bulk scalar fields in the strictly planar limit. In particular, after discussing the essential ingredients that goes into the derivation of the free field in subsection \ref{sec:klrev}, we extend the formalism to recover the \emph{interacting, local} (to cubic order) bulk scalar in this method. Then in section \ref{sec:metric}, we show how to use this construction to evaluate the metric everywhere in the bulk.\footnote{To avoid clutter, we will not always mention the phrase ``up to a conformal factor'', but this is always implied.} These last two sections are the main parts of our results. In particular, in subsection \ref{sec:cftvac}, we evaluate the bulk metric and show that we recover the AdS Poincar\'{e} patch. Next in subsection \ref{sec:fr} we point out that the field redefinition ambiguity in the work of \cite{Kabat:2017mun} has an important effect for our construction of bulk metric. For translational invariant states, the metric can be reproduced exactly in units of AdS radius i.e. when the conformal factor is a constant. However, for general states, this makes the metric recoverable up to a spacetime dependent, overall conformal factor. Modulo this subtlety, subsection \ref{sec:gAdS} and \ref{sec:tcft} are then respectively devoted towards evaluating the global bulk metrics for a CFT at zero and finite temperatures. In subsection \ref{sec:interiormetric}, we discuss how can this method be utilized to recover the bulk fields and consequently the metric, in the interior of a black hole or causal horizon. Finally, we conclude in section \ref{concl}, where we also discuss some possible directions for future studies.

\section{Recovering bulk fields using boundary modular Hamiltonian}\label{sec:DGrev}

In this section, we will briefly review the construction of \cite{Kabat:2017mun} which shows how from a CFT modular Hamiltonian associated with a given boundary subregion, can one construct a local bulk operator at any point inside the \emph{causal wedge} of the given subregion (see footnote \ref{deffoot} below for some definitions). The situation is simplest in AdS$_3$/CFT$_2$ case, so we will focus on this case in what follows. However, the whole set up can be generalized to arbitrary dimensions. We will also only discuss the time-independent situation. For the covariant generalization, see \cite{Kabat:2017mun}.

\begin{figure}[ht]
\begin{center}
\includegraphics[totalheight=0.3\textheight, angle=0]{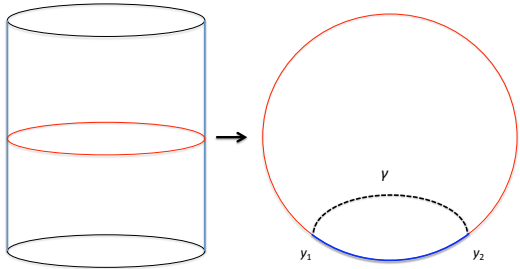}
\end{center}
\caption{Given a time slice in vacuum CFT (drawn on the left panel for pure AdS$_3$/CFT$_2$), denoted here by the red boundary circle, one can compute EE at the boundary using the following holographic RT prescription. As in the picture on the right panel, if we have a subregion of size $R$ (denoted here by end points $y_1$ and $y_2$ and in blue), then its entanglement with the rest of the CFT time slice is given by the area of the minimal area surface $\gamma$ (dashed line) in Planck units. \label{fig:RT}}
\end{figure}

We start with reviewing some basic ideas which, in hindsight, motivated this construction. Given a boundary subregion of size $R$ on a time-slice, the corresponding entanglement entropy (EE) at the boundary is holographically represented as the (one-quarter) area of a minimal area surface homologous to $R$ (known as Ryu-Takayanagi (RT) surface, $\gamma$) in Planck units \cite{Ryu:2006bv}. We depict the situation in figure \ref{fig:RT}. Computing the EE directly in field theory is usually a very hard problem. It is given by 
\be\label{eq:rt}
S_{EE}=-\text{Tr}\left(\rho_R\log \rho_R\right)=\frac{\text{Area}(\gamma)}{4G_N},
\ee
where 
\[
\rho_R=e^{-H_{mod}}
\]
is the reduced density matrix for the subregion $R$ and $H_{mod}$ is the associated modular Hamiltonian. The second relation in \eqref{eq:rt} comes from the RT proposal, which, as we can see, reduces the vastly complicated problem of taking logarithms of a density matrix to a relatively much simpler problem of computing area of a geometric surface in some manifold. The modular Hamiltonian or entanglement Hamiltonian, is in general a very complicated, non-local operator, but for spherical entangling surfaces, and especially for CFT$_2$ are quite easy to write down. To show how it works, let's consider the Poincar\'{e} patch of AdS$_3$,\footnote{We will denote the AdS radius by $l$, and reserve the symbol $R$ for the subregion size on the boundary.}
\begin{equation}
ds^2=\frac{l^2}{Z^2}(-dT^2+dZ^2+dX^2),
\end{equation}
with the dual CFT supported on Minkowski space,
\begin{equation}
ds^2=-dT^2+dX^2.
\end{equation} 
Introducing lightcone coordinates 
\begin{equation}
\xi=X-T\qquad\text{and}\qquad  \bar{\xi}=X+T,
\end{equation}
one can define the domain of dependence, $D$, of the boundary subregion as the region bounded by future and past directed intersecting light rays starting from the end points of the subregion.\footnote{For future reference, we point out that the bulk region which is comprised of the intersection of the bulk causal future and causal past of $D$ is called the causal wedge (CW) of the region $A$. On the other hand, let's call the bulk region which is bounded by the RT surface and $A$ as $\mathcal{R}_A$, i.e.
 \[
\text{boundary of}\; \mathcal{R}_A=\partial \mathcal{R}_A=A\cup\gamma.
\]
Then, the bulk domain of dependence of $\mathcal{R}_A$ is known as the entanglement wedge (EW). In simple situations like ours, CW=EW. However, they usually satisfy EW$\supseteq$CW. See e.g. \cite{Headrick:2014cta,Almheiri:2014lwa}.\label{deffoot}} Its upper tip $y^{\mu}$ and lower tip $x^{\mu}$ can be described using the light-front coordinates
\begin{equation}
\label{uv}
(u,\bar{u})=(x^1-x^0,x^1+x^0)\qquad\text{and}\qquad (v,\bar{v})=(y^1-y^0, y^1+y^0).
\end{equation}
If the end points of the subregion are $y_1$ and $y_2$ then
\begin{equation}
u=y_2, \ \ v=y_1, \ \  \bar{u}=y_1, \ \ \bar{v}=y_2\,.
\end{equation}
In these coordinates the CFT vacuum modular Hamiltonian is written as
\cite{Casini:2011kv}
\begin{equation}
H_{mod}=H_{mod}^{(R)}+H_{mod}^{(L)}=2\pi\int_{v}^{u}d\xi \frac{(u-\xi)(\xi-v)}{u-v}T_{\xi\xi}(\xi)+2\pi \int_{\bar{u}}^{\bar{v}}d\bar{\xi} \frac{(\bar{v}-\bar{\xi})(\bar{\xi}-\bar{u})}{\bar{v}-\bar{u}}\bar{T}_{\bar{\xi}\bar{\xi}}(\bar{\xi}).
\label{hmod}
\end{equation}
Here $T_{\xi\xi}$ and so on are the CFT$_2$ stress tensors. It can then be shown that a free bulk scalar $\Phi^{(0)}$ of mass $m$, when integrated over the minimal area RT surface mentioned above, commutes with the boundary modular Hamiltonian. That is\footnote{Below $C_{bulk}$ is a bulk normalization factor and $G_{N}$ is the Newton's constant. We have also dropped the zero from the superscript of $\Phi$ to avoid clutter. It signified that the bulk field is free, which will always be the case throughout our discussions with the exception of subsection \ref{subsec:pert}, where we will introduce the next order correction $\Phi^{(1)}$ to free fields $\Phi^{(0)}$ which should be added for the scalar field to be local at the cubic order of bulk perturbation.}
\be\label{eq:phiHcommute}
[Q,H_{mod}]=0 \qquad\text{where}\qquad Q = \frac{C_{bulk}}{8 \pi G_N} \int_\gamma ds \, \Phi.
\ee
This is most easily obtained by writing down the bulk field $\Phi$ as a smeared CFT operator $\mathcal{O}_{\Delta}$ \`{a} la HKLL and then using the commutation relations between the boundary stress tensors and the operator $\mathcal{O}_\Delta$ \cite{Hamilton:2005ju,Hamilton:2006az,Hamilton:2006fh}.

\subsection{CFT constraints for bulk locality}\label{sec:klrev}

\begin{figure}[ht]
\begin{center}
\includegraphics[totalheight=0.3\textheight, angle=0]{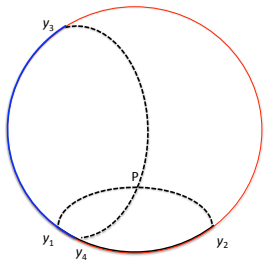}
\end{center}
\caption{Shown here are two overlapping boundary subregions on the CFT$_2$ time slice (red circle). The two subregions have endpoints $[y_1,y_2]$ and $[y_3,y_4]$ (black and blue arcs respectively) and their overlapping region is from $y_1$ to $y_4$. As a result, they intersect at point $P$. From the results of the previous subsection, a bulk field at point $P$ must commute with the modular Hamiltonians of both these regions and is hence localized at $P$. \label{fig:KLmodham}}
\end{figure}

Observations such as \eqref{eq:phiHcommute} play a crucial role in writing down a local bulk field in AdS. In fact, this result is intuitively obvious, the RT surface acts as a bifurcation surface for the flow under a modular Hamiltonian \cite{Jafferis:2015del}. It is a generalization to what happens for a Rindler wedge.\footnote{In AdS, a Rindler wedge is a bulk causal wedge corresponding to a subregion which extends to half of the AdS boundary.} For a Rindler wedge, the corresponding modular Hamiltonian is a boost generator which doesn't change the origin of the Rindler horizon \cite{Harlow:2014yka}.

However, one needs to work with a slightly modified version of \eqref{eq:phiHcommute} in order to construct a bulk local operator. The point is, even though a bulk scalar $\Phi$ commutes with the modular Hamiltonian when it is located on the RT surface, it doesn't necessarily imply the opposite. Namely, there is an infinite set of bulk operators residing on the bulk region complimentary to $\mathcal{R}_A$ (of the bulk time slice of figure \ref{fig:RT}), which also commute with $H_{mod}$. In order to tackle this, \cite{Kabat:2017mun} defined an extended modular Hamiltonian given by
\be
\tilde{H}_{mod}=\tilde{H}_{mod}^{(R)}+\tilde{H}_{mod}^{(L)}
\ee
where,
\begin{eqnarray}
\nonumber \tilde{H}_{mod}^{(R)}&=&2\pi \int_{-\infty}^{\infty} \frac{(w-y_1)(y_2-w)}{y_2-y_1} T_{ww}(w)\\
\tilde{H}_{mod}^{(L)}&=&2\pi \int_{-\infty}^{\infty} \frac{(\bar{w}-y_1)(y_2-\bar{w})}{y_2-y_1} T_{\bar{w}\bar{w}}(\bar{w}).
\end{eqnarray}
So one just extends the range of integrations compared to the regular modular Hamiltonian defined in \eqref{hmod}. As a result, its action within the boundary subregion is same as the one for the usual boundary modular Hamiltonian. In fact, it can be explicitly shown that \cite{Kabat:2017mun}
\begin{equation}\label{eq:pres}
[\tilde{H}_{mod},\Phi(Z,X,T=0)]=0,
\end{equation}
provided that
\begin{equation}\label{eq:onrt}
Z^2-(y_1+y_2)X+y_1y_2+X^2=0\,.
\end{equation}
If the metric is AdS$_3$, \eqref{eq:onrt} is simply the condition that the bulk point $(Z,X,T=0)$ lies on a spacelike geodesic whose endpoints hit the boundary at $(T=0,y_1)$ and $(T=0,y_2)$. 

However, \eqref{eq:pres} and \eqref{eq:onrt} are much more powerful than \eqref{eq:phiHcommute}. The extended modular Hamiltonian is given by 
\be
\tilde{H}_{mod}=H_{mod,A}-H_{mod,A^c},
\ee
where $H_{mod,A}=H_{mod}$ and $H_{mod,A^c}$ are respectively the modular Hamiltonians for the subregion $A$ and its complimentary boundary region $A^c$. So, as we have shown in figure \ref{fig:KLmodham}, the commutation relation in \eqref{eq:pres} and the associated equation \eqref{eq:onrt} clearly suggests that two intersecting subregions naturally select\footnote{This is true for AdS$_3$. For AdS$_{d+1}$, we need $d$ intersecting regions to pin-point a localized bulk operator.} a spatially localized bulk field at point $P$ which commutes with the extended modular Hamiltonians of both the regions. These commutation conditions then serve as  algebraic constraints on the `bulk' fields $\Phi$, where the bulk fields are written as
\begin{equation}\label{eq:bbansatz}
\Phi(X)=\int dt' dy' g(p,q){\cal O}(q,p).
\end{equation}
At this stage \eqref{eq:bbansatz} is simply an ansatz and $g(p,q)$ are the `smearing functions' that we want to ultimately derive. Here $q=X-t'+iy'$ and $p=X+t'+iy'$ and $X$ is currently a free variable. Of course, the above complexification of bulk spatial coordinates from $X$ to $X+iy'$ is motivated from earlier results of HKLL \cite{Hamilton:2006az,Hamilton:2006fh}, but at this stage, this is an educated guess. 

In what follows, we will represent this extended modular Hamiltonians for regions bounded by $[y_1,y_2]$ and $[y_3,y_4]$ by $\tilde{H}^{12}_{mod}$ and $\tilde{H}^{34}_{mod}$ and so on. These above ansatz and definitions, along with
\begin{equation}
(y_2-y_1)[\tilde{H}_{mod}^{12},\Phi(\xi, \bar{\xi})]=0 \qquad\text{and}\qquad (y_4-y_3)[\tilde{H}_{mod}^{34},\Phi(\xi, \bar{\xi})]=0
\label{eq:modhamconstraint}
\end{equation}
give us the correct smearing function with its support over the boundary points given by the intersection of spacelike lightcone from the point $P$ and the boundary. More precisely, one finds the corresponding smearing function $g(p,q)=K(t,x,z|t',y')$ to be given by the $d=2$ version of \cite{Kabat:2017mun}\footnote{In our notation, for $d=2$, $\vec{x},z,t$ will be replaced by uppercase $X,Z,T$.}
\begin{align}\label{eq:poincaresmear}
\Phi^{(0)}(t,\vec{x},z)&=c_\Delta\int dt'd^{d-1}\vec{y}'K(t,\vec{x},z|t',\vec{y}')\mathcal{O}_\Delta(t+t',\vec{x}+i\vec{y}')\nonumber\\
&=\frac{\Gamma\left(\Delta-\frac{d}{2}+1\right)}{\pi^{d/2}\Gamma\left(\Delta-d+1\right)}\int_{t'^2+|\vec{y}'|^2<z^2}dt'd^{d-1}\vec{y}'\left(\frac{z^2-t'^2-y'^2}{z}\right)^{\Delta-d}\mathcal{O}_\Delta(t+t',\vec{x}+i\vec{y}')\,.\nonumber\\
\end{align}
This is precisely the HKLL expression \cite{Hamilton:2005ju,Hamilton:2006az,Hamilton:2006fh}. However, now it is an outcome of a purely CFT calculations without using any bulk metric or bulk equations of motion. 

\subsection{Perturbative locality using modular Hamiltonians}\label{subsec:pert}

However, it turns out that the above-mentioned modular Hamiltonian constraints are far more powerful and can be utilized to obtain the subleading $1/N$ corrections to bulk locality. This can be expected, as the modular Hamiltonian that satisfy the constraints \eqref{eq:modhamconstraint} can be written purely using the CFT$_2$ conformal group generators in the following way:
\begin{equation}
	\tilde{H}_{mod}^{12}=\frac{2\pi}{(y_2-y_1)}\left[Q_0+y_1y_2P_0+(y_1+y_2)M_{01}\right]\, ,
\end{equation}
with 
\[
Q_0=i(\bar{L}_1-L_1),\qquad P_0=i(\bar{L}_{-1}-L_{-1})\qquad \text{and}\qquad M_{01}=i(\bar{L}_0-L_0).
\]
Thus, at this point, it is surprising that the resulting solutions for $\Phi$ in \eqref{eq:modhamconstraint} are only free fields $\Phi^{(0)}$ in the bulk, as in \eqref{eq:poincaresmear}.\footnote{We thank Gilad Lifschytz for discussions on this point.} Indeed, it turns out that this is not true, and the resulting set of solutions can at least be extended, so that one also recovers CFT fields which mimic interacting (to first order) bulk scalars.

The simplest way to see this is to try to solve the modular constraints \eqref{eq:modhamconstraint} with an ansatz general than \eqref{eq:bbansatz}. In particular, let us consider an ansatz 
\begin{equation}
	\Phi(X)=\int dt' dy' g(p,q){\cal O}_{\Delta}(q,p)+\sum_{n}a_n\int dt' dy' g_n(p,q){\cal O}_{\Delta_n}(q,p)\,,
\end{equation}
with the only difference being an infinite sum of smeared conformal \emph{primaries} with dimensions $\Delta_n$ and associated coefficients $a_n$ which are $\mathcal{O}(1/N)$ suppressed with respect to the leading counterpart.

If we plug this ansatz in \eqref{eq:modhamconstraint}, then we get an additional term in the constraint condition, similar to what we had before, but which contain infinite number of terms. Namely, we have
\begin{equation}\label{eq:pertconstr}
	(y_2-y_1)\left[\tilde{H}_{mod}^{12},\int dt' dy' g(p,q){\cal O}_{\Delta}(q,p)\right]+\sum_{n}a_n(y_2-y_1)\left[\tilde{H}_{mod}^{12},\int dt' dy' g_n(p,q){\cal O}_{\Delta_n}(q,p)\right]=0\,,
\end{equation}
and similarly for $\tilde{H}_{mod}^{34}$. 

In general, it is hopeless to try to solve for all the unknown functions $g(p,q)$ and $g_n(p,q)$'s from just two sets of equations. However, the derivation of 
\begin{equation}
	g(p,q)=c_{\Delta}\left[Z^2-(p-X_0)(q-X_0)\right]^{\Delta-2},\qquad\text{with}\qquad X_0=\frac{y_1y_2-y_3y_4}{y_1+y_2-y_3-y_4}
\end{equation}
and subsequently of \eqref{eq:poincaresmear}, is independent of the choice of $\Delta$ and only relies on the fact that the boundary operator $\mathcal{O}_{\Delta}$ is a conformal primary \cite{Kabat:2017mun}. 
Hence there is at least one set of solutions for \eqref{eq:pertconstr}, which gives the same functions $g_{n}(p,q)$'s as $g(p,q)$ with the only difference being the dependence on $\Delta$. In other words, a consistent set of solutions for \eqref{eq:pertconstr} is given by 
\begin{equation}
	g(p,q)=c_{\Delta}\left[Z^2-(p-X_0)(q-X_0)\right]^{\Delta-2}\qquad \text{and}\qquad g_n(p,q)=d_{\Delta_n}\left[Z^2-(p-X_0)(q-X_0)\right]^{\Delta_n-2}.
\end{equation}
Above, $d_{\Delta_n}$ is an overall dimension dependent factor which is undetermined along with the coefficients $a_n$. Therefore, the solution for the bulk field now boils down to 
\begin{align}\label{eq:pertresgen}
	\Phi(t,\vec{x},z)&=c_\Delta\int dt'd^{d-1}\vec{y}'K(t,\vec{x},z|t',\vec{y}')\mathcal{O}_\Delta(t+t',\vec{x}+i\vec{y}')\nonumber\\
	&+\sum_{n}a_nd_{\Delta_n}\int dt'd^{d-1}\vec{y}'K_{\Delta_n}(t,\vec{x},z|t',\vec{y}')\mathcal{O}_{\Delta_n}(t+t',\vec{x}+i\vec{y}')
\end{align}

Using bulk microcausality and the free field smearing function, the second term above already looks like the correction $\Phi^{(1)}$ that one needs to add to the free bulk field in order to recover bulk locality at subleading order of $\mathcal{O}(1/N)$ \cite{Kabat:2011rz}. However, without the knowledge of the bulk metric, it is not possible to argue for bulk lightcone singularity structures and consequently bulk microcausality. However, what we show in the next section is that the free bulk field $\Phi^{(0)}$ is actually \emph{sufficient} to determine the bulk metric up to an overall conformal factor, which in turn, \emph{completely specifies} the bulk microcausality structures. So, we can use the constraints from bulk microcausality subsequently, to compute what the required correction $\Phi^{(1)}$ is.

\section{Bulk metric from CFT data}\label{sec:metric}

In the previous subsection \ref{sec:klrev}, we briefly reviewed how \cite{Kabat:2017mun} extracted the smearing function appropriate for local bulk scalars in pure AdS$_3$ without knowledge of the bulk geometry (metric) or any bulk equation of motion. Since the end result is the same as those of \cite{Hamilton:2006az,Hamilton:2006fh}, it is guaranteed that these scalars will satisfy the correct bulk equations and microcausal commutation relations. In what follows we want to build on the above construction to extract the bulk metric structure purely from the boundary subregion data of modular Hamiltonians and CFT correlators.

\begin{figure}[ht]
\begin{center}
\includegraphics[totalheight=0.3\textheight, angle=0]{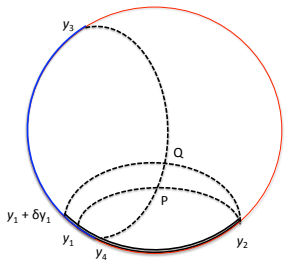}
\end{center}
\caption{We have perturbed one of the subregions (from $y_1$ to $y_2$) by a small amount, so that the new subregion is from $y_1+\delta y_1$ to $y_2$. It then gives us a different intersection point $Q$ where the bulk field is localized. $P$ and $Q$ are of course, spacelike separated. \label{fig:pert_region}}
\end{figure}

As a by-product of the construction of the previous section, one finds that the correct smearing function is obtained only if the $Z$ appearing in \eqref{eq:poincaresmear} above is given by (for $d=2$)
\be
Z^2=(y_1+y_2)X_{0}-y_1y_2-X_{0}^{2},
\label{eq:intsectcond}
\ee
where
\be
X_0=\frac{y_1y_2-y_3y_4}{y_1+y_2-y_3-y_4}.
\ee
This is of course what is expected, since $X_0,Z$ above are nothing but the solution of the two minimal area surfaces with endpoints $y_i$ and $y_{i+1}$ ($i=1,3$), namely
\be\label{eq:minsurf}
(X-y_i)(y_{i+1}-X)=Z^2.
\ee
That is, they give the coordinates of the intersection point $P$ in figure \ref{fig:KLmodham}.\footnote{We now see how the emergent radial direction $Z$ in the bulk comes about. In this framework, it simply coordinatizes the location of the field $\Phi$ in terms of the boundary coordinates as in \eqref{eq:intsectcond}. It makes the bulk, one higher dimensional as compared to the CFT.} However, we can think of shifting one of the end points of one of the subregions slightly (figure \ref{fig:pert_region}) and look for the resulting intersection point. For example, if we only shift $y_1$ to $y_1+\delta y_1$, then the resulting intersection point $Q$ has new coordinates $\tilde{X}_0$ and $\tilde{Z}$ given by\footnote{The quantities to start with are symmetric under interchanges of $y_1\leftrightarrow y_2$ and $y_3\leftrightarrow y_4$. So, if e.g. we also perturb the other end point from $y_2$ to $y_2+\delta y_2$, there will be an additional order $\delta y_2$ term, which is same as the one written above, but $y_1$ and $y_2$ interchanged.}
\begin{align}\label{newintersect}
&\tilde{X}_0=X_0+\frac{y_n}{y_c^2}\delta y_1-\frac{y_n}{y_c^3}(\delta y_1)^2+\dots\qquad\text{and}\nonumber\\
&\tilde{Z}=Z\left[1+\frac{2y_s-y_c(y_3+y_4)}{2(y_1-y_3)(y_1-y_4)y_c}\delta y_1+\dots\right],
\end{align}
where
\be\label{eq:ydefns}
y_n=(y_2-y_3)(y_2-y_4),\qquad y_c=y_1+y_2-y_3-y_4\qquad\text{and}\qquad y_s=y_1y_2-y_3y_4.
\ee
Now as this CFT prescription gives us the correct AdS$_3$ smearing function, it is guaranteed that the two point correlation function of two such smeared CFT fields $\Phi$ (starting from two-point function of CFT operators $\mathcal{O}$) will furnish the correct form for AdS$_3$ supergravity (SUGRA) correlators,
\begin{align}
\langle\Phi(x_1,z_1)\Phi(x_2,z_2)\rangle_{CFT}&=\int\int K(x_1,z_1|x)K(x_2,z_2|x')\langle\mathcal{O}(x)\mathcal{O}(x')\rangle_{CFT}\nonumber\\
&=G_{SUGRA} (x_1,z_1;x_2,z_2)\,,
\end{align}
just as it did for \cite{Hamilton:2006fh}. However, the crucial point to remember is that it is now an end-result of a CFT calculation which doesn't rely on any bulk dynamics.

So, we can write down the resulting bulk two-point function for two $\Phi$ fields located at $P$ and $Q$ of figure \ref{fig:pert_region} and the corresponding answer is guaranteed to be \cite{Lowe:2009mq,Hamilton:2006fh,Hamilton:2007wj}
\be\label{eq:cftres}
\langle\Phi(X_0,Z)\Phi(\tilde{X}_0,\tilde{Z})\rangle_{bulk}=c_\Delta\frac{1}{\sqrt{\sigma^2-1}}\frac{1}{(\sigma+\sqrt{\sigma^2-1})^{\Delta-1}}.
\ee
Here $c_\Delta$ is a constant factor dependent on CFT parameters\footnote{This is not exactly true, as there is a subtle issue of field redefinitions and their effects on $c_{\Delta}$. This is often overlooked in literature and we have discussed it below in subsection \ref{sec:fr}. We should also point out that the scalars on the left side of \eqref{eq:cftres} does not have the correct canonical bulk dimension. For example, in order to recover the correct bulk correlator, one needs to take into account factors of AdS radius.} and 
\be\label{eq:sigmadef}
\sigma=\frac{Z^2+\tilde{Z}^2+(X_0-\tilde{X}_0)^2}{2Z\tilde{Z}}.
\ee
We have written it suggestively in this format as for AdS, $\sigma$ is the AdS invariant distance between two points $X_0,Z$ and $\tilde{X}_0,\tilde{Z}$. Note that we have denoted the subscript of the bulk correlator above as just `bulk', as we do not yet know what its metric is and this is what we want to find out.

\subsection{Expanding Green's function}

Now to extract the bulk metric we will have to equate the \emph{CFT result} \eqref{eq:cftres} with the most general expression of bulk 2-point correlators involving scalars (of some mass $m=m(\Delta)$), written in terms of the geodesic distance in a generally curved spacetime (endowed with a metric $g_{MN}$ which is hitherto unknown) in some limit, such as the WKB (large mass) where bulk supergravity correlators are expected to have some universal form. Let's first reparametrize the bulk correlators in terms of a function $L$ such that\footnote{$l_{bulk}$ is a dimensionful length scale which simply exists because we know that we have an emergent radial direction.} 
\begin{equation}\label{eq:sigmadrel}
\sigma=\cosh\frac{L}{l_{bulk}},
\end{equation}
using which, we have
\begin{equation}\label{eq:terminprop}
\frac{1}{(\sigma+\sqrt{\sigma^2-1})^{\Delta-1}}=e^{-\frac{L}{l_{bulk}}(\Delta-1)}.
\end{equation}
On the other hand, 
\begin{equation}\label{eq:beforelimit}
(\sigma^2-1)=\frac{1}{4}e^{2L/l_{bulk}}\left(1+e^{-4L/l_{bulk}}\right)-\frac{1}{2}.
\end{equation}

%

At this point, we can take various limits. To begin with, here we will be interested in $\Delta\to \infty$ limit (keeping $L$ fixed) which signifies the stationary phase or WKB type approximation. In this case, from \eqref{eq:terminprop}, we see that the leading behavior of the bulk-bulk propagator becomes
\begin{equation}\label{eq:wkblim}
\langle\Phi(X_0,Z)\Phi(\tilde{X}_0,\tilde{Z})\rangle_{bulk}\Big{|}_{\Delta\gg 1}\approx \frac{c_{\Delta}}{\sqrt{\sigma^2-1}} e^{-\frac{L\Delta}{l_{bulk}}}=c'_{\Delta}e^{-\frac{L\Delta}{l_{bulk}}}.
\end{equation}


\subsection{Metric for Poincar\'{e} AdS}\label{sec:cftvac}

\eqref{eq:wkblim} is quite well-known as an expression of bulk propagator in general spacetimes at the leading order of geodesic distance \cite{Birrell:1982ix,Louko:2000tp} and evidently $L$ is a geodesic distance in the bulk. Thus our reparametrized expression can be used to extract the bulk metric structure. In fact, comparison of our \eqref{eq:wkblim} with the general results of \cite{Louko:2000tp} already tells us that the metric is AdS, as it is a well-known expression for asymptotically AdS spacetimes. However, we can make it more quantitative by writing ($M,N$ denote bulk indices)
\be\label{eq:Ldef}
L=\int\sqrt{g_{MN}dx^{M}dx^{N}},
\ee
and then computing the metric using \eqref{eq:sigmadef} and \eqref{eq:sigmadrel}. For example, in general dimensions, the metric component $g_{ZZ}$ or $g_{XX}$ can be extracted by choosing the two points to have same $X$ and $T$ or same $Z$ and $T$ coordinates respectively. The calculation of $g_{TT}$ will also follow the same procedure. 

Let's show the extraction of $g_{ZZ}$ in detail. Using WKB limit, we have already recognized the quantity $L$ defined through \eqref{eq:sigmadrel} as the geodesic distance. But now, we will instead consider the limit where the two bulk points are infinitesimally close, as we discussed back in figure \ref{fig:pert_region}. This, along with \eqref{eq:sigmadef}, then give us 
\begin{align*}
\sigma\Big{|}_{X_0\to \tilde{X}_0} &\approx 1+\frac{1}{2}\frac{dZ^2}{l_{bulk}^2}g_{ZZ}
\end{align*}
which yields,
\[
g_{ZZ}=\frac{l_{bulk}^2}{Z^2}.
\]
Note that this way we can extract the local value of the metric arbitrarily deep into the bulk and similarly we can fix different coordinates to extract out $g_{TT}$ and $g_{XX}$. The final result is of course $|g_{TT}|=g_{XX}=g_{ZZ}=\frac{l_{bulk}^2}{Z^2}$.

Note that our choice of $\sigma$ back in \eqref{eq:sigmadef} already helps us understand that the bulk metric is diagonal. In other words, after extracting the diagonal components as above, we can try to solve for off-diagonal components for two closely separated points. This yields (here $M\neq N$),
\[
\frac{dZ^2+dX^2-dT^2}{2Z^2}\approx\left(\frac{g_{MM}}{l_{bulk}^2}dx_M^2+\frac{g_{MN}}{l_{bulk}^2}dx_Mdx_N\right).
\]
If we now take two of the coordinates to be different and the last one to be same, this clearly indicates that the off-diagonal elements vanish. 

\subsection{Field redefinition freedom}\label{sec:fr}

At this point, it is important to note a key feature of this construction. As the subregion information of $y_i$'s are directly embedded in the bulk field's smearing formula \eqref{eq:poincaresmear} and subsequently in the two-point function \eqref{eq:cftres}, it is clear that the freedom of choosing an infinite pairs of subregions in order to obtain a given bulk point is also inharent in such a formalism. Moreover, the CFT constraints \eqref{eq:modhamconstraint} are also satisfied if the bulk fields are multiplied with some arbitrary classical function of spacetime. This later freedom was already noted in \cite{Kabat:2017mun}, which makes the coefficients such as $c_\Delta$ appearing in \eqref{eq:poincaresmear} position dependent. This is precisely the bulk field redefinition property ($x$ collectively denotes all bulk coordinates)
\be\label{eq:fr}
\Phi(x)\to f(x)\Phi(x)
\ee
of the above-mentioned smearing prescription. But, if we only consider translational invariant states at the boundary, from the perspective of  duality, the bulk states are also expected to have the same invariance. This reduces the possible sets of field redefinitions, in which $f$ can be taken as a function of $z$ alone. In fact, using the correct normalization of bulk 1-point function as $z\to 0$, it is easy to realize that the corresponding field redefinitions are just constant scaling of the bulk scalar \cite{Kabat:2017mun}.\footnote{Within HKLL, field redefinitions were also considered at the perturbative orders of $1/N$ \cite{Kabat:2015swa}.} In what follows, we will consider both situations where boundary translational symmetry may or may not be broken and study its effects on metric extraction. As noted in \cite{Kabat:2017mun}, these redefinition ambiguities keep the bulk singularity structures intact and hence by extension, do not affect microcausal prescriptions in higher orders of perturbation in $1/N$ (which only make use of the bulk singularity structures). In particular, we will point out that it makes the evaluation of the metric possible up to an overall conformal factor (See \cite{Engelhardt:2016wgb} for a similar conclusion but independent bulk arguments). Below we will make these statements quantitative.

Let's first consider the situation with full generality, i.e. when the CFT state doesn't have any translation invariance. In this case, we can write the field redefinitions acting as
\be
\langle\Phi(x)\Phi(\tilde{x})\rangle_{CFT}\to \tilde{f}(x, L) \langle\Phi(x)\Phi(\tilde{x})\rangle_{SUGRA}.
\ee

In order to understand the effects of field redefinitions as in \eqref{eq:fr}, we again turn to our bulk two-point correlator as obtained within the CFT \eqref{eq:cftres}. We start by studying the large $\Delta$ limit of this CFT correlation function which has been noted in \eqref{eq:wkblim}. Under \eqref{eq:fr}, the correlator \eqref{eq:wkblim} changes to\footnote{to avoid clutter, for the rest of this subsection we will put $l_{bulk}=1$ in the exponent of the WKB correlator \eqref{eq:wkblim}. We can always restore it by looking at the scaling dimension of the exponent.}
\be\label{eq:fr_wkblim}
\langle\Phi(x)\Phi(\tilde{x})\rangle_{CFT}=f_1(x)f_2(\tilde{x})e^{-\Delta L}/\sqrt{\sigma^2-1}=e^{F-\Delta L}/\sqrt{\sigma^2-1},\qquad \text{where}\qquad F=\log{f_1f_2}. 
\ee
In order to carry out the steps illustrated in subsection \ref{sec:cftvac} above, we must be able to identify $L$ as a geodesic distance. But now we see that due to an arbitrary function $F$, the required exponential behavior may be tampered. So, we will discuss various choices of $F$ below and see what are their effects on the identification of $L$.

Note that at the stage of \eqref{eq:fr_wkblim}, we are in the limit where the product, $\Delta L\to \infty$ and fixed $L$. To this end we treat three separate cases which cover all possible behaviors of $F$
\be\label{eq:Fexp}
F \sim \sum_{n\geq 0}c_n(x)(\Delta L)^n,\hspace{4mm} F \sim \sum_{l\geq1}c_l(x)\log^{l} (\Delta L),\hspace{4mm}  F \sim \sum_{m<0} c_m(x)(\Delta L)^m.
\ee
Here $n\geq 0$, $m<0$ and $c_n,c_l$ and $c_m$'s can be arbitrary functions of the bulk coordinates $x$. In order to sort out the physical field redefinitions, we assume that we know the correct singularity structures of CFT correlators (including their proper normalization) and we also expect that bulk lightcone singularities will show up in appropriate limits. We also do not expect spurious singularities to show up in unexpected situations, e.g. when $L\to \infty$ limit is taken in a suitable manner. 

\begin{enumerate}
\item The first case with $n>1$ is harmless as far as the identification of $L$ goes. They all come with a higher power of $L$ in the two-point correlator, whereas from \cite{Birrell:1982ix,Louko:2000tp}, we know that the geodesic distance always comes as a linear power of $\Delta L$ in the exponential.

Thus for field redefinitions of this type, we do not have any ambiguity in extracting the bulk metric. However, the ambiguity arise for the case of $n=0,1$, which we will discuss only at the end.

\item In a similar way, we note that for non-zero $c_l(x)$, $e^F$ is a polynomial in $\Delta L$. So once again, from the distinctive exponential behavior we can always identify $L$ as the geodesic distance. For example, we can take a derivative with respect to $\Delta$ in  \eqref{eq:fr_wkblim} and subsequently take an $L\to 0$ limit. This erases any polynomial contribution that may be present after the field redefinition and distinguishes the exponential behavior.

\item The $c_m(x)$'s are of course subleading terms in the WKB limit and do not contribute to the two-point function at our required order.

\item\label{frsource} We now revisit first case with $n=0,1$. Indeed a field redefinition of type $F=c_1(x)\Delta L+c_0(x)$ is consistent with the bulk singularity structure in the $\Delta\to\infty$ limit (also when $c_1,c_0$'s are constants).\footnote{The terms involving $c_0$ is of course also subleading with respect to the term involving $c_1$.} But this precisely makes the definition of $L$ ambiguous by an $x$-dependent factor, which in turn induces a $x$-\emph{dependent} conformal factor in the evaluation of the bulk metric using \eqref{eq:sigmadrel},\eqref{eq:Ldef}.

We also note that it is not possible to get rid of such an ambiguity by just demanding the boundary behavior of the bulk correlators. For example, $f_1f_2$ could be $f_1f_2=Ne^{cz^n\Delta L}$, such that as $z\to 0$, we recover the correct CFT two-point function with the required normalization factor $N$. These types of field redefintions, which still induces a conformal factor, are then invisible in the $z\to 0$ limit.

\end{enumerate}

To end the section, we briefly mention about the special case of translationally invariant states, which is simply obtained from the results above for general state. As mentioned before, in this case, we can consider the field redefintions as multiplication by constant factors. Thus, from the point \ref{frsource} discussed above, now the metric is obtained up to a constant rescaling of the conformal factor. This makes complete sense, as this is precisely our CFT ignorance about the precise value of $l_{bulk}$. In other words, for translationally invariant states, we can compute the metric exactly in the units of AdS radius $l_{bulk}$.

Now that we have obtained the bulk conformal metric, we can finally fully reconstruct the local bulk fields at the subleading order of boundary perturbations. Fortunately, the knowledge of the conformal metric is enough to understand bulk lightcone singularity structures and consequently that of bulk microcausality. Thus, we can get back to the expression we obtained at the end of subsection \ref{subsec:pert}, namely \eqref{eq:pertresgen}, and demand that the resulting bulk field satisfies bulk microcausality. As shown in \cite{Kabat:2011rz,Kabat:2015swa}, this is now sufficient to select the suitable conformal (double-trace) primaries that one requires to use in the infinite sum of \eqref{eq:pertresgen}.\footnote{However, We should note that the gauge dressings of scalars, which are important at subleading order in $1/N$ might also show up in certain gauges alongside the $\Delta L$ factor in the exponent of \eqref{eq:wkblim}. It seems that choosing e.g. the holographic gauge \cite{Kabat:2012hp}, it is possible to still extract the correct geodesic length $L$, even in that case.}

\subsection{Extracting the metric for global AdS}\label{sec:gAdS}

In what follows, we will keep aside the issues of the field redefinition and carry on the `observer dependent' construction for various other coordinate systems. Because we have now represented the required formulas of subsection \ref{sec:cftvac} in terms of AdS covariant quantities, our proof is valid for any other coordinate patches such as global AdS or Rindler patch. However, the situation is different for BTZ BH as we have a physical singularity there, even though the spacetime is locally AdS$_3$. We will turn to these cases in the next subsections, but for now we deal with the next simplest example after the Poincar\'{e} AdS$_3$ patch, which is the global AdS$_3$ case. In this scenario, if we define the function\footnote{Here we have replaced the nomenclature of $Z,X,T$ of the previous section by $\tau,\rho$ and $\Omega$ and now the computation of \cite{Kabat:2017mun} needs to be performed in these coordinate systems. The fact that we will obtain the correct smearing function and hence the correct bulk fields, are guaranteed due to the fact that the resulting expression of the smearing function in \cite{Kabat:2017mun} is fully covariant.}
\be\label{eq:globalgeodesic}
\sigma(x|x')=\frac{\cos(\tau-\tau')-\sin\rho\sin\rho'\cos(\Omega-\Omega')}{\cos\rho\cos\rho'},
\ee
then the above mentioned CFT techniques give us the bulk to bulk Green's function \eqref{eq:cftres}. We can then proceed like in the last section and define a quantity $L$ via the relation \eqref{eq:sigmadrel}. Once again, as soon as we take the WKB limit, it enables us to interpret $L$ as the geodesic distance. 

In particular, using e.g. $\Omega\to \Omega'$ and $\rho\to \rho'$, we get 
\be
\sigma\big{|}_{\stackrel{\Omega\to \Omega'}{\rho\to \rho'}}\approx 1+\frac{d\tau^2}{2l_{bulk}^2}g_{\tau\tau}\approx 1-\frac{d\tau^2}{2\cosh^2\rho}.
\ee
Similarly, taking $\Omega\to \Omega'$ and $\tau\to \tau'$ and then $\tau\to \tau'$ and $\rho\to \rho'$, we can respectively obtain 
\[
g_{\rho\rho}=\frac{l_{bulk}^2}{\cos^2\rho}\qquad \text{and}\qquad g_{\Omega\Omega}=\frac{l_{bulk}^2\sin^2\rho}{\cos^2\rho}.
\]
Once more, the off-diagonal terms of the metric vanish. Thus we recover the global AdS metric 
\be\label{eq:gads}
ds^2=\frac{l_{bulk}^2}{\cos^2\rho}(-d\tau^2+d\rho^2+\sin^2\rho d\Omega_{d-1}^2),
\ee
as expected.

\subsection{CFT at finite temperature}\label{sec:tcft}

So far, we have only considered the CFT vacuum state and its subregions in different boundary coordinate patches. But the above arguments and procedures already tell us that the calculation will go through similarly if we make a coordinate change to go to the Rindler patch of the AdS spacetime. This enables us to define a natural Rindler state at the boundary which corresponds to an observer experiencing a thermal bath. Ultimately the Rindler state can be upgraded to another boundary CFT state at a finite temperature, which corresponds to the presence of a (topological) BTZ black hole in the bulk \cite{Banados:1992wn}. In subsection \ref{sec:btzout} we will discuss how to extract the exterior metric to the BTZ geometry, but first, in order to understand the subtleties involving the BTZ case, we will warm up with the example of extracting the Rindler patch.

\subsubsection{Bulk metric outside the Rindler horizon}\label{sec:rindout}

Once again, we start with the CFT procedure of constructing bulk fields in terms of intersecting RT surfaces. As has been already pointed out in \cite{Kabat:2017mun}, this procedure gives rise to the correct bulk operator in the background of Rindler AdS. For a bulk field in the (e.g.) right exterior of a Rindler spacetime, it is obtained as (the coordinate conventions are given below in \eqref{eq:rindcoord})
\be
\label{RightRindler}
\phi(\htt,r,\hp) = c_{\Delta}(r_+) \int_{\hbox{\small spacelike}}
\hspace{-0.5cm} dx dy \,
\lim_{r' \rightarrow \infty} (\sigma / r')^{\Delta - 2} \mathcal{O}^{Rindler,R}(\htt + x, \hp + i y),
\ee
where as $r' \rightarrow \infty$, the AdS invariant distance becomes 
\be\label{eq:ads3rindcovdist2}
\sigma(\htt,r,\hp \vert \htt + x,r',\hp + iy) = {rr' \over r_+^2} \left[\cos y -
\left(1 - {r_+^2 \over r^2}\right)^{1/2} \cosh x \right].
\ee
This is precisely what one obtains in the old HKLL procedure \cite{Hamilton:2006fh}. Thus it is guaranteed that this CFT prescription will give us the necessary bulk to bulk two-point function. However, if we only use a definite spectrum of a given CFT (to which a given Rindler observer has an access to) with their corresponding modular Hamiltonians, the resulting bulk operator and hence the bulk metric will turn out to be located \emph{outside} the Rindler horizon.\footnote{\label{fnint}This is not true in general. Indeed, for generic spacetimes, the RT surface supported on one CFT can enter the causal, event horizon \cite{Headrick:2014cta}. Because our construction relies on RT surfaces, rather than causal wedge, in those cases, a sub-horizon metric construction is in principle possible.} We will briefly visit the question of interior (of Rindler horizon) operator construction in section \ref{sec:interiormetric}.

A simple way to extract the exterior Rindler metric is to redo the whole calculation by defining \cite{Hamilton:2006fh,Kabat:2017mun}
\be\label{eq:ads3rindcovdist}
\sigma(\htt,r,\hp \vert \htt',r',\hp') = {rr' \over r_+^2} \cosh (\hp-\hp') -
\left({r^2 \over r_+^2}-1\right)^{1/2}\left({r'^2 \over r_+^2}-1\right)^{1/2} \cosh (\htt-\htt').
\ee
Because this gives the correct bulk scalar \emph{outside the right Rindler horizon}, the corresponding two-point function is again of the form \eqref{eq:cftres}. Clearly the two bulk points here are located at $(\htt,r,\hp)$ and $(\htt + x,r',\hp + iy)$ respectively. From here, we can extract the Rindler metric, but we can only access the outside part of the Rindler horizon.

In this case, the geodesic distance $L$ is once again obtained from \eqref{eq:sigmadrel}. In fact, the easiest way to understand that the procedure will go through for exterior (to Rindler horizon) operators in Rindler and the above choice of $\sigma$ in \eqref{eq:ads3rindcovdist} is the correct choice, it's best to understand Rindler AdS as a coordinate transformed version from the global patch \eqref{eq:gads}.\footnote{In \cite{Hamilton:2006fh} the Rindler smearing function was obtained from the Poincar\'{e} patch by going to dS via an analytic continuation and then using the bulk equations.} The Rindler metric is given by\footnote{The coordinates $\htt,\hp$ are related to $t,\phi$ by
\be\label{eq:hatnhatreln}
\htt=\frac{r_+t}{l_{bulk}^2}\qquad\text{and}\qquad \hp=\frac{r_+\phi}{l_{bulk}}.
\ee}
\be\label{eq:rindcoord}
ds^2=-\frac{r^2-r_+^2}{l_{bulk}^2}dt^2+\frac{l_{bulk}^2}{r^2-r_+^2}dr^2+r^2d\phi^2,
\ee
which are related to the global coordinates by the following coordinate transformations (see e.g. \cite{Mann:1996ze}):
\[
i\tau=\frac{r_+}{l_{bulk}}\phi,\qquad\Omega=\frac{r_+}{l_{bulk}^2}(it)\qquad\text{and}\qquad \cos\rho=\frac{r_+}{r}.
\]
Indeed, the geodesic distance in Rindler \eqref{eq:ads3rindcovdist} can be obtained from the geodesic distance in global AdS \eqref{eq:globalgeodesic}, by using the above coordinate transformations. In fact, the covariant relations \eqref{eq:sigmadrel} and the related constructions of the metric (outside the Rindler horizon) go through unchanged.

But we can also extract the metric quite straight-forwardly. Using \eqref{eq:ads3rindcovdist} and \eqref{eq:hatnhatreln} in \eqref{eq:sigmadrel}, we find (using the expression of geodesic distance in terms of the metric and looking at diagonal components as before)
\be
g_{rr}=\frac{l_{bulk}^2}{(r^2-r_+^2)},\qquad g_{\phi\phi}=r^2\qquad\text{and}\qquad g_{tt}=-\frac{r^2-r_+^2}{l_{bulk}^2}.
\ee
This is precisely the  Rindler coordinate as shown in \eqref{eq:rindcoord}.\footnote{We must mention some subtle issues with the field redefiniton here, which is also applicable for the case of BTZ black holes discussed afterwards. Even though the state is translation invariant, the argument for constant field redefinitions is now a bit more involved. This is because, using the new temperature scale $\beta^{-1}$, one can expand the $c_{\Delta}(r_+)$ in \eqref{RightRindler} in a series such as
\[
c_{\Delta}(r_+)=1+\dots+c_n\left(\frac{z}{\beta}\right)^n+\dots.
\]
It then requires a bit more work to show that $c_{\Delta}(r_+)$ is a given function of $\Delta, d$ and $r_+$. However, since Rindler is pure AdS and the left hand side of \eqref{RightRindler} is a scalar, the right hand side must reduce to the right hand side of the global AdS scalar upon making the Rindler to global coordinate transformations. This will again lead to $c_\Delta$ being a pure function of $\Delta, r_+$ and $d$.}

\subsubsection{Bulk metric outside the BTZ horizon}\label{sec:btzout}

Once we are done with Rindler, there's not much to be done for the BTZ metric's exterior. In fact, the BTZ geometry is obtained from the Rindler-AdS geometry by taking the range of $\phi$ coordinate between 0 and $2\pi$, so locally they are indistinguishable \cite{Banados:1992wn,Hamilton:2006fh}. The key ingredient that we need to show is that using the $\sigma$ as defined in \eqref{eq:ads3rindcovdist}, the WKB connection between the BTZ correlator and the BTZ geodesic distance goes through. 

In other words, we know that for Rindler-AdS (RAdS) and in the WKB approximation, the following is true:
\[
\langle\Phi(\htt,r,\hp)\Phi(\htt',r'\hp')\rangle_{AdS}=c'_{\Delta}e^{-\Delta L_{RAdS}/l_{bulk}}.
\]
Here we require that for BTZ, a similar relation holds in the WKB approximation as well:
\be\label{eq:btzwkb}
\langle\Phi(\htt,r,\hp)\Phi(\htt',r'\hp')\rangle_{BTZ}=c'_{\Delta}e^{-\Delta L_{BTZ}/l_{bulk}}
\ee
for $\sigma$ still given by \eqref{eq:ads3rindcovdist}. Of course, given this, the derivation of the metric is exactly similar as before (the only difference being the above-mentioned periodicity of $\phi$). Writing $L_{BTZ}$ in terms of BTZ metric components, we recover the BTZ metric.

The way to see that \eqref{eq:btzwkb} is correct is to note that the BTZ correlator is given by the image sum of AdS correlators \cite{Lifschytz:1993eb}:
\begin{align}\label{eq:btzcorr}
\langle\Phi(\htt,r,\hp)\Phi(\htt',r'\hp')\rangle_{BTZ}&=\sum_{n=-\infty}^{\infty}\langle\Phi(\htt,r,\hp)\Phi(\htt',r'\hp'+2\pi n)\rangle_{AdS}\nonumber\\
&=\sum_{n=-\infty}^{\infty}c_\Delta\frac{1}{\sqrt{\sigma_n^2-1}}\frac{1}{(\sigma_n+\sqrt{\sigma_n^2-1})^{\Delta-1}},
\end{align}
where $\sigma_n$ are given by  
\[
\sigma_n={rr' \over r_+^2} \cosh (\hp-\hp'-2\pi n) -
\left({r^2 \over r_+^2}-1\right)^{1/2}\left({r'^2 \over r_+^2}-1\right)^{1/2} \cosh (\htt-\htt').
\]
However, when we consider WKB approximation, i.e. $\Delta\to\infty$, we see that the leading order (in $\Delta$) contribution in the right hand side of \eqref{eq:btzcorr} comes from $\sigma_{n=0}$, which is nothing but $\sigma$ defined in \eqref{eq:ads3rindcovdist}. All the $\sigma_n$ terms for $n\neq 0$ (both positive and negative) are much larger and negligible when taken to a very high power as in $\frac{1}{(\sigma_n+\sqrt{\sigma_n^2-1})^{\Delta-1}}$. 

\subsection{Bulk fields and metric inside the Rindler horizon}\label{sec:interiormetric}

Finally, we make some brief comments regarding a possible interior operator construction in the modular Hamiltonian approach, in particular, for Rindler/ thermofield double (TFD) states. This question is at the very  heart of the black hole firewall problem \cite{Almheiri:2013hfa} and is still open. Although our lack of knowledge about modular Hamiltonians for generic states renders such construction more subtle, its entanglement nature gives us a hope to go beyond the causal wedge. For example, for Rindler, one can simply recast the results of global or Poincar\'{e} AdS in Rindler coordinates, which clearly states that there are no physical problems behind such a construction. But on the other hand, if we want to stick to the subregions that any given Rindler observer has access to, we can't simply use the global AdS result.

In the bulk approach of constructing the local operators \cite{Hamilton:2006fh}, the problem appeared because as soon as one extends the bulk operator beyond the Rindler horizon (let's say for the right Rindler exterior), the corresponding smearing region goes beyond the right Rindler boundary region. This compelled them to use the anti-podal map, which brings the extra extended region back to the left side of the Rindler wedge and finally another analytic continuation in boundary time to bring it to the right.

In our case however, the problem arises because there are no two subregions on the same side of a Rindler patch (say right side), for which the corresponding RT surface goes beyond the Rindler horizon and intersects at a point (see however footnote \ref{fnint}). In such scenarios, we can consider two intersecting RT surfaces stretching between two sides of the full Rindler geometry, but located on the same time slice \cite{Hartman:2013qma}. This guarantees that they intersect at a single point inside Rindler horizon. However, little is known about the field theory counterpart of such an entanglement entropy, let alone its associated modular Hamiltonian. In fact, this problem is a particular example of a more general question of whether we can generically construct local bulk observables inside the EW of a given boundary subregion, in terms of simple boundary operators located in the subregion. As mentioned before, EW is usually bigger than the causal wedge and such a `simple' construction is no longer possible \cite{Dong:2016eik,Faulkner:2017vdd}.

%

\section{Discussions and outlook}\label{concl}

In this work, we have provided a recipe to extract the bulk metric using the local (scalar) operator construction from modular Hamiltonian data of boundary subregions \cite{Kabat:2017mun}. We showed that, the knowledge of CFT data, associated with boundary modular Hamiltonian can indeed reproduce the bulk metric, but up to a conformal factor. For the CFT states we considered, this conformal factor turned out to be a constant. This is completely expected for such a pure CFT reconstruction, as CFT is blind to the AdS length scale and this ambiguity doesn't do anything short of what's expected for a well-posed duality such as AdS/CFT. In order to reproduce the bulk metric exactly, one requires some more information on the bulk-boundary connection, such as the Ryu-Takayanagi prescription for entanglement entropy mentioned before. This was e.g. the approach of \cite{Faulkner:2013ica,Faulkner:2017tkh}. In fact, a by-product of the modular Hamiltonian approach was that the fields $\Phi$ satisfied \eqref{eq:modhamconstraint}, only if their radial location was given by \eqref{eq:onrt}. So, alternatively, we can consider \eqref{eq:onrt} to be the geodesic equation in the bulk, as from the point of view of the bulk modular flow (which is dual to the boundary modular flow), it is the only bulk curve for which the bulk field doesn't modular transform. Such an additional equation/information regarding bulk then precisely computes the unknown conformal factor, thereby computing the bulk metric \emph{exactly}.

However in this work, we have solely focused on special states of the CFT, such as the vacuum or thermal states. These states have a very high degree of symmetry and as a consequence the modular Hamiltonian is easy to construct. Also we worked with a two dimensional boundary, for which the RT surface is one-dimensional. So here we make some comments when each of these conditions are generalized. First we discuss the case of higher dimensions, i.e. AdS$_{d+1}$/CFT$_{d}$. We expect the higher dimension case to be a straightforward, though geometrically more involved construction. RT surfaces being co-dimension 2 surfaces, will intersect on a co-dimension 3 surface. Thus one will need $d$ number of RT surfaces to intersect, so as to single out a point in the $(d+1)$ dimensional bulk, and further to solve $d$ algebraic equations to reproduce the HKLL construction (and in turn to compute the metric). We leave this higher dimensional construction for future work.

Second, we discuss the situation regarding the CFT states. In this paper, we looked at vacuum state and the thermal state (thermofield double or TFD states) and we derived the bulk metric which is either pure AdS$_{3}$ or the Rindler wedge of AdS$_{3}$ or the BTZ. Although this is nice and reassuring to see that the recipe works, these bulk metrics are all too familiar to us and this work should be regarded more as a proof of principle. To realize the full potential of our approach, one needs to extend our results to determine metrics for states of the CFT for which the bulk metric is unknown. The main issue here is to find expressions of Modular Hamiltonians for CFT states perturbatively close to the vacuum or thermal states. See e.g. \cite{Sarosi:2017rsq} (and references therein) for some progress in this direction.  Although these states are perturbatively excited with respect to the CFT vacuum, these are not to be conflated with perturbative excitations in $1/N$. The perturbatively excited CFT states we are interested are the ones which cause a perturbation of the bulk metric. On the other hand, CFT states which are perturbative in $1/N$ can be thought of as interacting multi-particle states on the unbackreacted bulk metric. Incidentally, the $1/N$ correction to the alternative HKLL construction \cite{Kabat:2017mun} is also an interesting issue. 

There are several other avenues for future work. One obvious direction is to look at the reconstruction of bulk gauge fields from boundary entanglement, and how bulk gauge redundancy is represented in the boundary modular Hamiltonian. But perhaps the most interesting question is to find an extension of the recipe to reconstruct local bulk fields and further the metric beyond the causal wedge, such as regions inside black hole horizons. For the two sided black holes, the construction should be a straightforward generalization of the Rindler horizon interior, as discussed in section \ref{sec:interiormetric}. Entanglement among subregions having support on both boundaries of the two Rindler wedges/exterior regions code the interior local operator or metric data. However for single sided black holes, the region inside the horizon are outside the causal wedge of the maximal boundary subregion. So unless the RT surface somehow extends beyond the horizon, the present recipe needs to be generalized for these cases in order to reconstruct regions beyond the causal wedge.

%

\vspace{1.5 cm}
\centerline{\bf Acknowledgements}

We thank Dan Kabat for his perceptive comments, especially regarding the field redefinition ambiguity in \cite{Kabat:2017mun}, and for giving his valuable feedback on the draft. SR also thanks the Albert Einstein Center for Fundamental Physics of the University of Bern for their hospitality, during which the project was conceived. DS thanks City College of New York, KITP and Columbia university for hospitalities during work on this project. The work of SR is supported by IIT Hyderabad seed grant SG/IITH/F171/2016-17/SG-47 and his travel to Bern was supported by IIT Block Grant. The work of DS is funded by the NCCR SwissMAP (The Mathematics of Physics) of the Swiss Science Foundation.
%

\providecommand{\href}[2]{#2}\begingroup\raggedright


\begin{thebibliography}{99}

\bibitem{Kabat:2017mun} 
  D.~Kabat and G.~Lifschytz,
  ``Local bulk physics from intersecting modular Hamiltonians,''
  JHEP {\bf 1706}, 120 (2017)
  doi:10.1007/JHEP06(2017)120
  [arXiv:1703.06523 [hep-th]].

\bibitem{Maldacena:1997re} 
  J.~M.~Maldacena,
  ``The Large N limit of superconformal field theories and supergravity,''
  Adv.\ Theor.\ Math.\ Phys.\  {\bf 2}, 231 (1998)
  [hep-th/9711200].
  
  \bibitem{Gubser:1998bc} 
  S.~S.~Gubser, I.~R.~Klebanov and A.~M.~Polyakov,
  ``Gauge theory correlators from noncritical string theory,''
  Phys.\ Lett.\ B {\bf 428}, 105 (1998)
  [hep-th/9802109].
  
  \bibitem{Witten:1998qj} 
  E.~Witten,
  ``Anti-de Sitter space and holography,''
  Adv.\ Theor.\ Math.\ Phys.\  {\bf 2}, 253 (1998)
  [hep-th/9802150].
  
   \bibitem{Hamilton:2005ju} 
  A.~Hamilton, D.~N.~Kabat, G.~Lifschytz and D.~A.~Lowe,
  ``Local bulk operators in AdS/CFT: A Boundary view of horizons and locality,''
  Phys.\ Rev.\ D {\bf 73}, 086003 (2006)
  doi:10.1103/PhysRevD.73.086003
  [hep-th/0506118].
  
  \bibitem{Hamilton:2006az} 
  A.~Hamilton, D.~N.~Kabat, G.~Lifschytz and D.~A.~Lowe,
  ``Holographic representation of local bulk operators,''
  Phys.\ Rev.\ D {\bf 74}, 066009 (2006)
  doi:10.1103/PhysRevD.74.066009
  [hep-th/0606141].
  
  \bibitem{Hamilton:2006fh} 
  A.~Hamilton, D.~N.~Kabat, G.~Lifschytz and D.~A.~Lowe,
  ``Local bulk operators in AdS/CFT: A Holographic description of the black hole interior,''
  Phys.\ Rev.\ D {\bf 75}, 106001 (2007)
  Erratum: [Phys.\ Rev.\ D {\bf 75}, 129902 (2007)]
  doi:10.1103/PhysRevD.75.106001, 10.1103/PhysRevD.75.129902
  [hep-th/0612053].
  
  \bibitem{Balasubramanian:1998sn} 
  V.~Balasubramanian, P.~Kraus and A.~E.~Lawrence,
  ``Bulk versus boundary dynamics in anti-de Sitter space-time,''
  Phys.\ Rev.\ D {\bf 59}, 046003 (1999)
  [hep-th/9805171].
  

  \bibitem{Banks:1998dd} 
  T.~Banks, M.~R.~Douglas, G.~T.~Horowitz and E.~J.~Martinec,
  ``AdS dynamics from conformal field theory,''
  hep-th/9808016.
  
  
  
   \bibitem{Kabat:2012hp} 
  D.~Kabat, G.~Lifschytz, S.~Roy and D.~Sarkar,
  ``Holographic representation of bulk fields with spin in AdS/CFT,''
  Phys.\ Rev.\ D {\bf 86}, 026004 (2012)
  [arXiv:1204.0126 [hep-th]].
  
\bibitem{Sarkar:2014dma} 
  D.~Sarkar and X.~Xiao,
  ``Holographic Representation of Higher Spin Gauge Fields,''
  Phys.\ Rev.\ D {\bf 91}, no. 8, 086004 (2015)
  doi:10.1103/PhysRevD.91.086004
  [arXiv:1411.4657 [hep-th]].
  
  \bibitem{Sarkar:2014jia} 
  D.~Sarkar,
  ``(A)dS holography with a cutoff,''
  Phys.\ Rev.\ D {\bf 90}, no. 8, 086005 (2014)
  doi:10.1103/PhysRevD.90.086005
  [arXiv:1408.0415 [hep-th]].
  
  \bibitem{Xiao:2014uea} 
  X.~Xiao,
  ``Holographic representation of local operators in de sitter space,''
  Phys.\ Rev.\ D {\bf 90}, no. 2, 024061 (2014)
  doi:10.1103/PhysRevD.90.024061
  [arXiv:1402.7080 [hep-th]].

  \bibitem{Lowe:2008ra} 
  D.~A.~Lowe and S.~Roy,
  ``Holographic description of asymptotically AdS(2) collapse geometries,''
  Phys.\ Rev.\ D {\bf 78}, 124017 (2008)
  doi:10.1103/PhysRevD.78.124017
  [arXiv:0810.1750 [hep-th]].
  
  \bibitem{Kabat:2014kfa} 
  D.~Kabat and G.~Lifschytz,
  ``Finite N and the failure of bulk locality: Black holes in AdS/CFT,''
  JHEP {\bf 1409}, 077 (2014)
  doi:10.1007/JHEP09(2014)077
  [arXiv:1405.6394 [hep-th]].
  
  \bibitem{Roy:2015pga} 
  S.~R.~Roy and D.~Sarkar,
  ``Hologram of a pure state black hole,''
  Phys.\ Rev.\ D {\bf 92}, 126003 (2015)
  doi:10.1103/PhysRevD.92.126003
  [arXiv:1505.03895 [hep-th]].
  
\bibitem{Kabat:2016rsx} 
  D.~Kabat and G.~Lifschytz,
  ``Asymmetric interiors for small black holes,''
  JHEP {\bf 1608}, 097 (2016)
  doi:10.1007/JHEP08(2016)097
  [arXiv:1601.05611 [hep-th]].  
  
  \bibitem{Roy:2017hcp} 
  S.~R.~Roy and D.~Sarkar,
  ``Holographic bulk reconstruction beyond (super)gravity,''
  Phys.\ Rev.\ D {\bf 96}, no. 8, 086018 (2017)
  doi:10.1103/PhysRevD.96.086018
  [arXiv:1704.06294 [hep-th]].
  
  \bibitem{Kabat:2011rz} 
  D.~Kabat, G.~Lifschytz and D.~A.~Lowe,
  ``Constructing local bulk observables in interacting AdS/CFT,''
  Phys.\ Rev.\ D {\bf 83}, 106009 (2011)
  [arXiv:1102.2910 [hep-th]].
  
  \bibitem{Kabat:2012av} 
  D.~Kabat and G.~Lifschytz,
  ``CFT representation of interacting bulk gauge fields in AdS,''
  Phys.\ Rev.\ D {\bf 87}, no. 8, 086004 (2013)
  [arXiv:1212.3788 [hep-th]].
  
  \bibitem{Kabat:2013wga} 
  D.~Kabat and G.~Lifschytz,
  ``Decoding the hologram: Scalar fields interacting with gravity,''
  Phys.\ Rev.\ D {\bf 89}, 066010 (2014)
  [arXiv:1311.3020 [hep-th]].

\bibitem{Kabat:2015swa} 
  D.~Kabat and G.~Lifschytz,
  ``Bulk equations of motion from CFT correlators,''
  JHEP {\bf 1509}, 059 (2015)
  doi:10.1007/JHEP09(2015)059
  [arXiv:1505.03755 [hep-th]].
  
  \bibitem{Kabat:2016zzr} 
  D.~Kabat and G.~Lifschytz,
  ``Locality, bulk equations of motion and the conformal bootstrap,''
  JHEP {\bf 1610}, 091 (2016)
  doi:10.1007/JHEP10(2016)091
  [arXiv:1603.06800 [hep-th]].
    
  \bibitem{Ryu:2006bv} 
  S.~Ryu and T.~Takayanagi,
  ``Holographic derivation of entanglement entropy from AdS/CFT,''
  Phys.\ Rev.\ Lett.\  {\bf 96}, 181602 (2006)
  doi:10.1103/PhysRevLett.96.181602
  [hep-th/0603001].
  
  \bibitem{VanRaamsdonk:2009ar} 
  M.~Van Raamsdonk,
  ``Comments on quantum gravity and entanglement,''
  arXiv:0907.2939 [hep-th].

  \bibitem{Faulkner:2013ica} 
  T.~Faulkner, M.~Guica, T.~Hartman, R.~C.~Myers and M.~Van Raamsdonk,
  ``Gravitation from Entanglement in Holographic CFTs,''
  JHEP {\bf 1403}, 051 (2014)
  doi:10.1007/JHEP03(2014)051
  [arXiv:1312.7856 [hep-th]].
    
    \bibitem{Almheiri:2014lwa} 
  A.~Almheiri, X.~Dong and D.~Harlow,
  ``Bulk Locality and Quantum Error Correction in AdS/CFT,''
  JHEP {\bf 1504}, 163 (2015)
  doi:10.1007/JHEP04(2015)163
  [arXiv:1411.7041 [hep-th]].
 
  \bibitem{Pastawski:2015qua} 
  F.~Pastawski, B.~Yoshida, D.~Harlow and J.~Preskill,
  ``Holographic quantum error-correcting codes: Toy models for the bulk/boundary correspondence,''
  JHEP {\bf 1506}, 149 (2015)
  doi:10.1007/JHEP06(2015)149
  [arXiv:1503.06237 [hep-th]].
    
  \bibitem{Jafferis:2015del} 
  D.~L.~Jafferis, A.~Lewkowycz, J.~Maldacena and S.~J.~Suh,
  ``Relative entropy equals bulk relative entropy,''
  JHEP {\bf 1606}, 004 (2016)
  doi:10.1007/JHEP06(2016)004
  [arXiv:1512.06431 [hep-th]].
  
  \bibitem{Sanches:2017xhn} 
  F.~Sanches and S.~J.~Weinberg,
  ``Boundary dual of bulk local operators,''
  Phys.\ Rev.\ D {\bf 96}, no. 2, 026004 (2017)
  doi:10.1103/PhysRevD.96.026004
  [arXiv:1703.07780 [hep-th]].
  
  \bibitem{Faulkner:2017tkh} 
  T.~Faulkner, F.~M.~Haehl, E.~Hijano, O.~Parrikar, C.~Rabideau and M.~Van Raamsdonk,
  ``Nonlinear Gravity from Entanglement in Conformal Field Theories,''
  JHEP {\bf 1708}, 057 (2017)
  doi:10.1007/JHEP08(2017)057
  [arXiv:1705.03026 [hep-th]].
  
\bibitem{Headrick:2014cta} 
  M.~Headrick, V.~E.~Hubeny, A.~Lawrence and M.~Rangamani,
  ``Causality \& holographic entanglement entropy,''
  JHEP {\bf 1412}, 162 (2014)
  doi:10.1007/JHEP12(2014)162
  [arXiv:1408.6300 [hep-th]].
  
  \bibitem{Casini:2011kv} 
  H.~Casini, M.~Huerta and R.~C.~Myers,
  ``Towards a derivation of holographic entanglement entropy,''
  JHEP {\bf 1105}, 036 (2011)
  doi:10.1007/JHEP05(2011)036
  [arXiv:1102.0440 [hep-th]].
  
  \bibitem{Harlow:2014yka} 
  D.~Harlow,
  ``Jerusalem Lectures on Black Holes and Quantum Information,''
  Rev.\ Mod.\ Phys.\  {\bf 88}, 015002 (2016)
  doi:10.1103/RevModPhys.88.015002
  [arXiv:1409.1231 [hep-th]].
  
  \bibitem{Lowe:2009mq} 
  D.~A.~Lowe,
  ``Black hole complementarity from AdS/CFT,''
  Phys.\ Rev.\ D {\bf 79}, 106008 (2009)
  doi:10.1103/PhysRevD.79.106008
  [arXiv:0903.1063 [hep-th]].
  
  \bibitem{Hamilton:2007wj} 
  A.~Hamilton, D.~N.~Kabat, G.~Lifschytz and D.~A.~Lowe,
  ``Local bulk operators in AdS/CFT and the fate of the BTZ singularity,''
  AMS/IP Stud.\ Adv.\ Math.\  {\bf 44}, 85 (2008)
  [arXiv:0710.4334 [hep-th]].
  
  \bibitem{Birrell:1982ix} 
  N.~D.~Birrell and P.~C.~W.~Davies,
  ``Quantum Fields in Curved Space,''
  doi:10.1017/CBO9780511622632
  
  \bibitem{Louko:2000tp} 
  J.~Louko, D.~Marolf and S.~F.~Ross,
  ``On geodesic propagators and black hole holography,''
  Phys.\ Rev.\ D {\bf 62}, 044041 (2000)
  doi:10.1103/PhysRevD.62.044041
  [hep-th/0002111].
  
  \bibitem{Engelhardt:2016wgb} 
  N.~Engelhardt and G.~T.~Horowitz,
  ``Towards a Reconstruction of General Bulk Metrics,''
  Class.\ Quant.\ Grav.\  {\bf 34}, no. 1, 015004 (2017)
  doi:10.1088/1361-6382/34/1/015004
  [arXiv:1605.01070 [hep-th]].
    
  \bibitem{Banados:1992wn} 
  M.~Banados, C.~Teitelboim and J.~Zanelli,
  ``The Black hole in three-dimensional space-time,''
  Phys.\ Rev.\ Lett.\  {\bf 69}, 1849 (1992)
  doi:10.1103/PhysRevLett.69.1849
  [hep-th/9204099].
  
  \bibitem{Mann:1996ze} 
  R.~B.~Mann and S.~N.~Solodukhin,
  ``Quantum scalar field on three-dimensional (BTZ) black hole instanton: Heat kernel, effective action and thermodynamics,''
  Phys.\ Rev.\ D {\bf 55}, 3622 (1997)
  doi:10.1103/PhysRevD.55.3622
  [hep-th/9609085].
  
  \bibitem{Lifschytz:1993eb} 
  G.~Lifschytz and M.~Ortiz,
  ``Scalar field quantization on the (2+1)-dimensional black hole background,''
  Phys.\ Rev.\ D {\bf 49}, 1929 (1994)
  doi:10.1103/PhysRevD.49.1929
  [gr-qc/9310008].
  
  \bibitem{Almheiri:2013hfa} 
  A.~Almheiri, D.~Marolf, J.~Polchinski, D.~Stanford and J.~Sully,
  ``An Apologia for Firewalls,''
  JHEP {\bf 1309}, 018 (2013)
  doi:10.1007/JHEP09(2013)018
  [arXiv:1304.6483 [hep-th]].
  
  \bibitem{Hartman:2013qma} 
  T.~Hartman and J.~Maldacena,
  ``Time Evolution of Entanglement Entropy from Black Hole Interiors,''
  JHEP {\bf 1305}, 014 (2013)
  doi:10.1007/JHEP05(2013)014
  [arXiv:1303.1080 [hep-th]].
  
  \bibitem{Dong:2016eik} 
  X.~Dong, D.~Harlow and A.~C.~Wall,
  ``Reconstruction of Bulk Operators within the Entanglement Wedge in Gauge-Gravity Duality,''
  Phys.\ Rev.\ Lett.\  {\bf 117}, no. 2, 021601 (2016)
  doi:10.1103/PhysRevLett.117.021601
  [arXiv:1601.05416 [hep-th]].
  
  \bibitem{Faulkner:2017vdd} 
  T.~Faulkner and A.~Lewkowycz,
  ``Bulk locality from modular flow,''
  JHEP {\bf 1707}, 151 (2017)
  doi:10.1007/JHEP07(2017)151
  [arXiv:1704.05464 [hep-th]].
  
\bibitem{Sarosi:2017rsq} 
  G.~Sárosi and T.~Ugajin,
  ``Modular Hamiltonians of excited states, OPE blocks and emergent bulk fields,''
  JHEP {\bf 1801}, 012 (2018)
  doi:10.1007/JHEP01(2018)012
  [arXiv:1705.01486 [hep-th]].
  
%
%
  
  


\end{thebibliography}
\end{document}